# Anomalous Hall Effect in $Sn_{1-x-y}Mn_xEu_yTe$ and $Sn_{1-x-y}Mn_xEr_yTe$ Mixed Crystals

K. Racka, I. Kuryliszyn, M. Arciszewska, W. Dobrowolski

*Institute of Physics, Polish Academy of Sciences, Warszawa, Poland*

J.-M. Broto, M. Goiran, O. Portugall, H. Rakoto, B. Raquet

*Laboratoire National des Champs Magnetiques Pulses, Toulouse, France*

V. Dugaev, E.I. Slynko, V.E. Slynko

*Chernivtsy Department of the Institute of Materials Science Problems, Ukraine*

Abstract

The Anomalous Hall Effect was investigated in IV-VI ferromagnetic semimagnetic semiconductors of $Sn_{1-x}Mn_xTe$ codoped with either Eu or Er. The analysis of experimental data: Hall resisitivity and magnetization showed that AHE coefficient $R_S$ depends on temperature, its value decreases with the temperature increase. We observe that above ferromagnet-paramagnet transition temperature $R_S$ changes sign. We discuss the possible physical mechanisms responsible for observed temperature dependence of $R_S$, particularly change of the sign.

## 1. INTRODUCTION

Anomalous Hall Effect (AHE), observed in ferromagnetic metals since the discovery of the Hall effect, have attracted recently much attention due to the rapidly growing interest in spin dependent transport phenomena [1]. In quest of materials for potential application in spintronics AHE was used in establishing an existence of a ferromagnetism in several semimagnetic (diluted magnetic) semiconductors. Studies of AHE provide also crucial information about physical mechanism responsible for ferromagnetism in these compounds [2].

The Hall resistivity of magnetic materials can be expressed as a sum of two terms – the normal part and anomalous one:

$$\rho_{Hall} = R_0 \cdot B + \mu_0 \cdot R_S \cdot M \tag{1}$$

where $R_0$ and $R_S$ are the normal and anomalous Hall coefficients, respectively, $B$ is the magnetic field, $\mu_0$ is the magnetic permeability and $M$ the magnetization. While the normal Hall effect results from the Lorenz force, the AHE in standard models applicable to metals is related with the asymmetric scattering of the charge carriers where the spin-orbit interaction plays the important role. Spin-orbit coupling prefers only distinguished directions what leads to the additional transversal current. This current gives the additional contribution to the Hall voltage. It is generally accepted that two mechanisms of scattering are responsible for the anomalous Hall effect: the skew-scattering (in presence of spin-orbit coupling, the amplitude of scattered wave packet becomes anisotropic) and the side-jump (the charge carrier changes its trajectory due to a lateral displacement) [3]. Recently a new approach to the theory of AHE in semimagnetic semiconductors was presented by Jungwirth et al. [4]. This theory relates the AHE to Berry phase acquired by a quasiparticle wave function upon traversing closed paths on the spin-split Fermi surface.

In this paper we present very preliminary results of AHE studies performed in IV-VI bulk semiconductor of $Sn_{1-x}Mn_xTe$ codoped with Eu or Er ions. We have chosen the IV-VI system for several reasons. These semiconductors crystallize in simple NaCl structure. Their energetic structure parameters are well known. It was established that in $Sn_{1-x}Mn_xTe$ and closely related $Pb_{1-x-y}Sn_xMn_yTe$ mixed crystals the RKKY interaction mediated via holes is responsible for variety of magnetic behaviors, in particular for ferromagnetic ordering at low temperatures [5]. Thus the paramagnet-ferromagnet phase transition temperature, $T_C$, may be controlled by changing not only Mn ions content but also free holes concentration (that can be tuned by thermal annealing or tin content variation). Concluding, we found $Sn_{1-x}Mn_xTe$ compounds ideal for AHE studies. We plan to investigate also the AHE in MBE grown epilayers of IV-VI ferromagnetic semiconductors, particularly the influence of interfaces on AHE in these compounds.

## 2. EXPERIMENT

Crystals used in the experiment were grown by the modified Bridgman method. Samples were cut perpendicularly to the ingot axis to reduce the effect of possible inhomogeneity. The chemical composition was determined using the X-ray dispersive fluorescence analysis ($x=0.01 \div 0.02$ and $y=0.06 \div 0.13$ for $Sn_{1-x-y}Eu_xMn_yTe$ samples and $x=0.001 \div 0.002$ and $y=0.11 \div 0.12$ for $Sn_{1-x-y}Er_xMn_yTe$ samples). Electrical contacts were made by use of indium solder and gold wire leads. The transport studies were performed in continuous flow cryostat using standard dc technique in the temperature range $2\ K \leq T \leq 300\ K$ and at magnetic fields $B$ up to 1.6 T. We have measured the Hall resistivity and conductivity as a function of magnetic field at various temperatures. The magnetic properties were studied by use of Lake Shore 7229 Magnetometer/Susceptometer. The magnetization as a function of magnetic field $M(B)$ was measured using extraction method up to 2 T at the same temperatures as for the transport investigations (the Hall resistivity and conductivity versus magnetic field).

The investigated $Sn_{1-x-y}Eu_xMn_yTe$ and $Sn_{1-x-y}Er_xMn_yTe$ crystals show p-type conductivity, the hole concentration occurred to be of the order of $10^{21}$ cm$^{-3}$. For all of the studied samples the paramagnet-ferromagnet phase transition is below 20 K (see Fig. 1). This enable us to perform measurements in a wide range of temperatures above (paramagnetic state) and below (ferromagnetic state) the Curie temperature $T_C$.

## 3. RESULTS AND DISCUSSION

Hall coefficient $R_H$ for the $Sn_{0.86}Eu_{0.01}Mn_{0.13}Te$ sample as a function of magnetic field for a few temperatures is presented on Fig. 2. Figure 3 presents experimental Hall resistivity $\rho_{Hall}$ vs. $B$ for the same sample at three various temperatures. We observe strongly nonlinear dependence of $\rho_{Hall}$ vs. $B$, the Hall resistivity reflects the magnetic field dependence of magnetization. Simultaneous analysis of both magnetization and transport data allows to extract the normal ($R_0$) and anomalous Hall effect ($R_S$) coefficient values. Taking the experimental evaluation of both $M(B)$ and $\rho_{Hall}(B)$ and using equation (1), the fitting procedure was applied to determine two parameters: $R_0$ and $R_S$. We have found that the anomalous Hall coefficient $R_S$ clearly depends on temperature and changes the sign (see Fig. 4 and Fig. 5 for Sn$_{1-x-y}$Eu$_x$Mn$_y$Te and for Sn$_{1-x-y}$Er$_x$Mn$_y$Te samples, respectively). The normal Hall coefficient $R_0$ seems to be practically temperature independent. Such behavior was not previously observed in similar materials [6].

The several physical mechanism can lead to the experimental results reported in this paper. First, we have analyzed the influence of the temperature on the energy spectrum parameters due to the temperature dependence of energy gap $E_G(T)$. Our calculations indicate that such mechanism is negligibly weak and does not explain the experimental results.

Second, we estimated the phonon scattering effect on temperature dependence of both components of conductivity: the anomalous Hall conductivity $\sigma_{AH}$ and usual conductivity $\sigma$. The theoretical analysis shows that $\sigma_{AH}$ is much more sensitive for temperature than $\sigma$, especially at low temperatures a slight decrease of $\sigma_{AH}$ with $T$ is predicted in agreement with our experiment. Nevertheless, the theoretical temperature dependence of $\sigma_{AH}$ is apparently weaker then determined experimentally. Moreover, the observed change of sign of $R_S(T)$ can not be understood in the frame of phonon scattering mechanism.

In general, two mechanisms of scattering: side-jump and skew-scattering are responsible for the anomalous Hall effect. In random alloys where the alloy composition is varied the interplay of these two mechanisms can lead to change of the sign of the AHE coefficient. Such effect may be also visible in the temperature dependence of $R_S$. Our estimations indicate that varying temperature, we modify the skew-scattering contribution and this may result in a change of AHE coefficient sign. This variation may results from the change of the phonon scattering induced by varying temperature.

Nonvanishing localization corrections to the $\sigma_{AH}$ within the skew-scattering mechanism may also lead to the change of sign for $R_S$.

Acknowledgements

This work was supported in part by the United States Army through its European Research Office Contract N68171-00-M-5929 and by European Community program ICA1-CT-2000-70018 (Center of Excellence CELDIS).

Figure captions

Fig. 1 Dependence of the magnetic susceptibility on temperature for $Sn_{1-x-y}Eu_xMn_yTe$ samples x=0.01, y=0.13 – $\Delta$; x=0.02, y=0.06 – $\nabla$) and for $Sn_{1-x-y}Er_xMn_yTe$ samples (x=0.001, y=0.11 – $\square$, x=0.002, y=0.12 – $\circ$)

Fig. 2 Hall coefficient $R_H$ for the $Sn_{0.86}Eu_{0.01}Mn_{0.13}Te$ sample as a function of magnetic field for a few temperatures: 7.5 K – ■, 13.4 K – $\nabla$, 18 K – $\circ$, 21.3 K – ●, 300 K – ▲.

Fig. 3 Hall resisitivity $\rho_{Hall}$ vs. magnetic field B for a few temperatures (circles). Sold line – result of fit to the formula $\rho_{Hall} = R_0 \cdot B + \mu_0 \cdot R_S \cdot M$.

Fig. 4 Dependence of the anomalous Hall coefficient $R_S$ on temperature for $Sn_{0.92}Eu_{0.02}Mn_{0.06}Te$.

Fig. 5 Dependence of the anomalous Hall coefficient $R_S$ on temperature for $Sn_{0.878}Er_{0.002}Mn_{0.12}Te$.

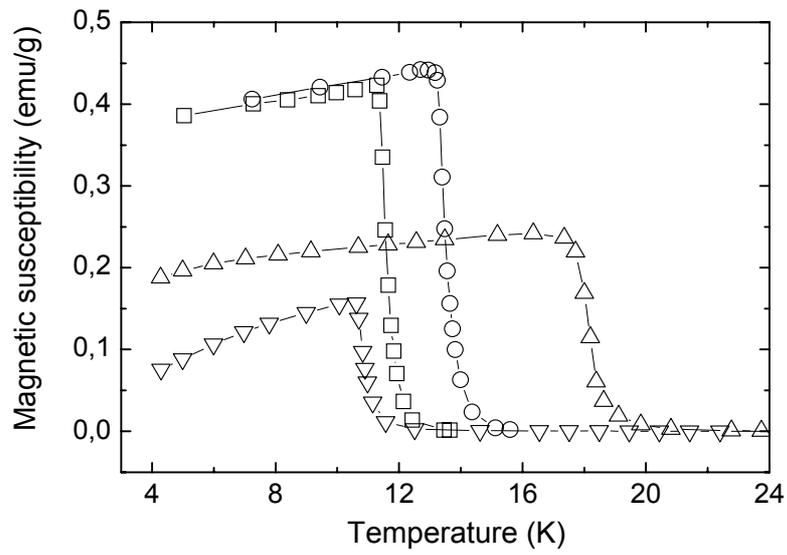

Fig. 1

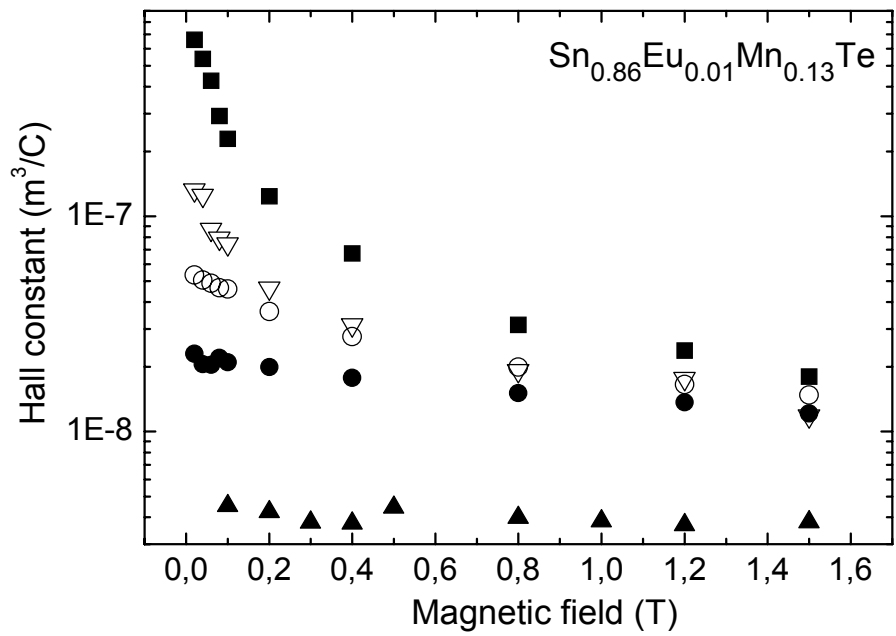

Fig. 2

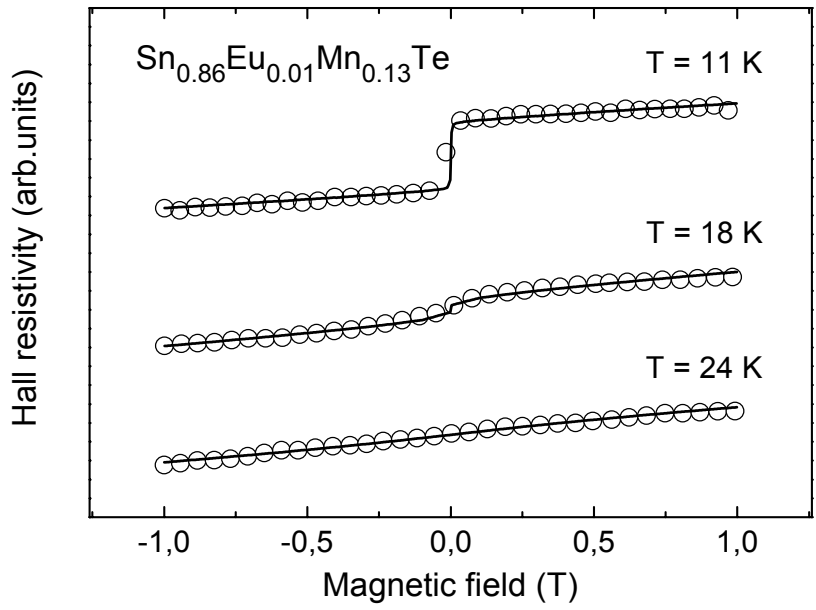

Fig. 3

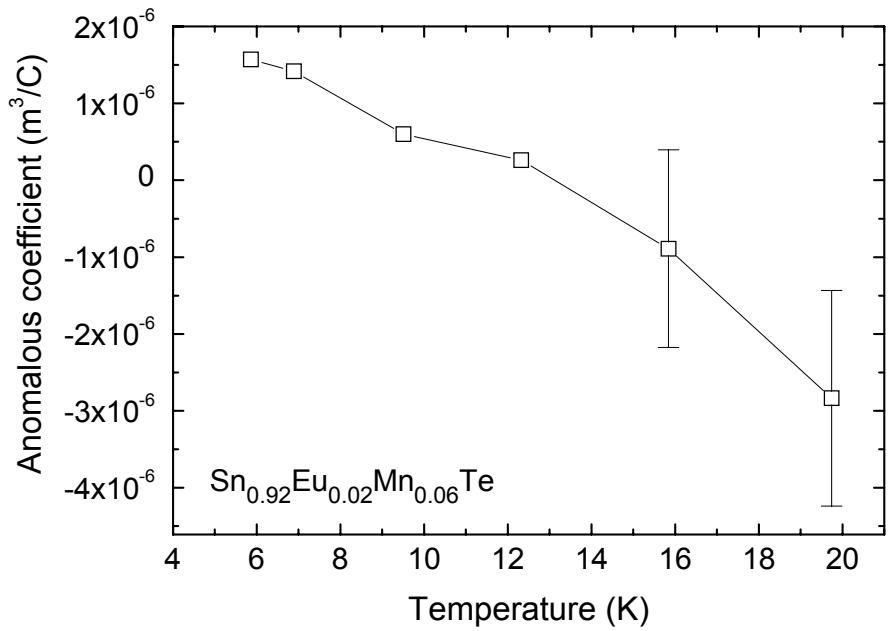

Fig. 4

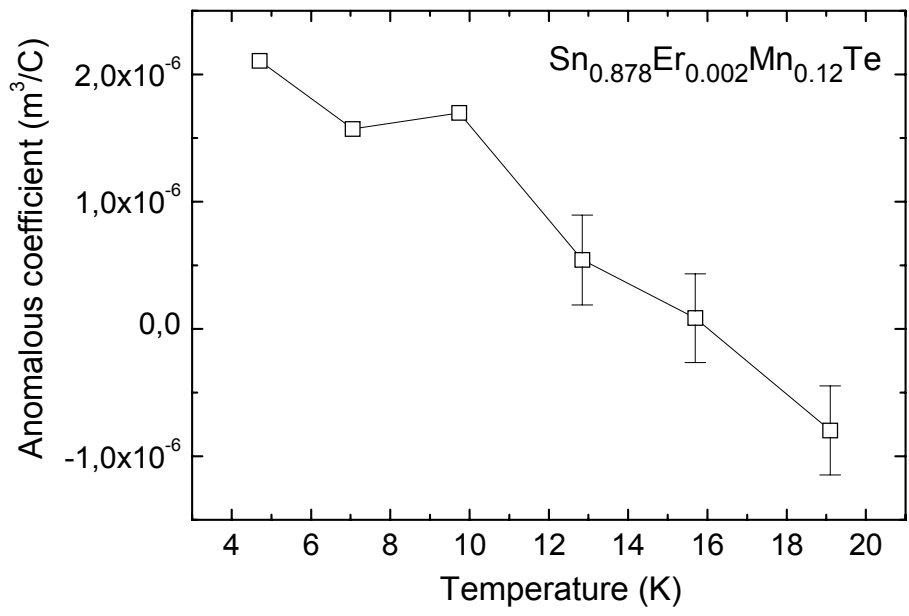

Fig. 5